\documentclass[12pt,a4paper]{article}
\usepackage[english]{babel}
\usepackage[T1]{fontenc}
\usepackage{ae,aecompl}
\usepackage[numbers,comma,square,compress]{natbib}
\usepackage{amsmath,amssymb,amsfonts}
\usepackage{bm}
\usepackage{ifpdf}
\ifpdf
  \usepackage[pdftex,a4paper,hmargin=25mm,vmargin=25mm]{geometry}
  \usepackage[pdftex,bookmarks=true]{hyperref}
\else
  \usepackage[dvips,a4paper,hmargin=25mm,vmargin=25mm]{geometry}
  \usepackage[dvips,bookmarks=true]{hyperref}
\fi

\numberwithin{equation}{section}

\newcommand{\email}[1]{Electronic address: \href{mailto:#1}{#1}}
\newcommand{\HL}{Ho\v{r}ava-Lifshitz }

\newcommand{\be}{\begin{equation}}
\newcommand{\ee}{\end{equation}}
\newcommand{\nn}{\nonumber}

\newcommand{\mc}[1]{\mathcal{#1}}
\newcommand{\mr}[1]{\mathrm{#1}}
\newcommand{\p}{\partial}
\newcommand{\bx}{\bm{x}}
\newcommand{\by}{\bm{y}}
\DeclareMathOperator{\rank}{rank}
\newcommand{\gM}[1][4]{{}^{(#1)}\!g}
\newcommand{\RM}[1][4]{{}^{(#1)}\!R}

\newcommand{\GM}[1][4]{{}^{(#1)}\!G}
\newcommand{\GMphi}[1][4]{{}^{(#1)}\!\mathcal{G}_{\phi}}
\newcommand{\projector}[2]{g^{#1}_{\phantom{#1}#2}}
\newcommand{\per}[1]{{}_\perp\!#1}
\newcommand{\perRM}[1][4]{{}^{(#1)}_{\,\,\perp}\!R}

\begin{document}

\begin{center}
{\Large Arnowitt-Deser-Misner representation and Hamiltonian\\
\vspace{.3em}
analysis of covariant renormalizable gravity}\\
\vspace{1em}
Masud Chaichian,\footnote{\email{masud.chaichian@helsinki.fi}}
Markku Oksanen,\footnote{\email{markku.oksanen@helsinki.fi}}
Anca Tureanu \footnote{\email{anca.tureanu@helsinki.fi}}\\
\vspace{1em}
\textit{Department of Physics, University of Helsinki, P.O. Box 64,\\
FI-00014
Helsinki, Finland}
\end{center}

\begin{abstract}
We study the recently proposed Covariant Renormalizable Gravity (CRG),
which aims to provide a generally covariant ultraviolet completion of
general relativity. We obtain a space-time decomposed form --- an
Arnowitt-Deser-Misner (ADM) representation --- of the CRG action. The
action is found to contain time derivatives of the gravitational fields
up to fourth order. Some ways to reduce the order of these time
derivatives are considered. The resulting action is analyzed using the
Hamiltonian formalism, which was originally adapted for constrained
theories by Dirac. It is shown that the theory has a consistent set of
constraints. It is, however, found that the theory exhibits four
propagating physical degrees of freedom. This is one degree of freedom
more than in Ho\v{r}ava-Lifshitz (HL) gravity and two more propagating
modes than in general relativity. One extra physical degree of freedom
has its origin in the higher order nature of the CRG action. The other
extra propagating mode is a consequence of a 
projectability condition similarly as in HL gravity. Some additional
gauge symmetry may need to be introduced in order to get rid of the
extra gravitational degrees of freedom.

\vspace{1.5em}
11.10.Ef (Lagrangian and Hamiltonian approach),
98.80.Cq (Particle-theory and field-theory models of the early Universe)
\end{abstract}

\section{Introduction}
In the recent years modified theories of gravity have attracted a
considerable amount of attention. These modifications of General
Relativity (GR) aim to improve the behaviour of the theory either at
high energies (renormalizability, early-time universe) or at large
distances (cosmology, galaxies), and sometimes at both regimes (see
\cite{Chaichian:2010a,Carloni:2010} for such an attempt), in order to
achieve a better agreement with observational data or a more plausible
theoretical framework. Recently the so-called \emph{Covariant
Renormalizable Gravity (CRG)} was proposed in Refs.
\cite{Nojiri:2010a,Nojiri:2010b,Nojiri:2010c}. For a review of CRG and
its comparison with other modified gravities, one can see
\cite{Nojiri:2010d}. CRG aims to provide a power-counting renormalizable
field theory of gravity that is covariant under spacetime diffeomorphism
and possesses local Lorentz invariance at the fundamental level. Lorentz
invariance of the graviton propagator of CRG is, however, broken
dynamically at high 
energies. This is achieved by introducing an exotic fluid of unknown
origin, which is coupled to spacetime in a rather complicated way. \HL
(HL) gravity \cite{Horava:2009uw} --- another power-counting
renormalizable field theory of gravity --- is based on the idea that
space and time scale differently at high energies,
\be
\bx \rightarrow b \bx \,,\qquad t \rightarrow b^z t \,,
\ee
with a dynamic critical exponent $z=1,2,3,\ldots\,$. This enables one to
modify the ultraviolet behaviour of the graviton propagator to
$|\bm{k}|^{-2z}$, where $\bm{k}$ is the spatial momentum. In $D$ spatial
dimensions, $z=D$ ensures the theory is power-counting renormalizable.
Such spacetime admits a preferred foliation into spatial hypersurfaces,
and hence the local Lorentz invariance is broken.
Note that more general theories of such sort have been proposed in Ref.
\cite{Carloni:2010}.
CRG achieves a similar ultraviolet behaviour of the graviton propagator
as HL gravity, but without introducing explicitly Lorentz noninvariant
terms into the action. 
However, we should note that the renormalizability of CRG, as well as of
the HL model, is assumed only based on the power-counting arguments.
There are several potential pathologies that could ruin the
renormalizability of this theory, such as gradient instabilities, ghosts
or strong coupling.
Since a violation of Lorentz invariance has never been observed, one
could argue that CRG is a more natural modification of GR than the
explicitly Lorentz noninvariant ones, in particular HL gravity and its
generalizations (see e.g.
\cite{Chaichian:2010a,Carloni:2010,Sotiriou:2009,Blas:2010}).

Hamiltonian formalism provides a powerful tool for the analysis of 
constrained systems such as gravity. For example, it has been shown that
HL gravity can be physically consistent ($N>0$) at high energies only if
the projectability condition is imposed on the lapse function $N$
\cite{Henneaux:2010}: $N=N(t)$. A similar result has also been obtained
for the more general modified $F(R)$ HL gravity \cite{Chaichian:2010b}.
In this paper we seek to understand CRG from the point of view of
Hamiltonian analysis.

First we obtain the Arnowitt-Deser-Misner (ADM) representations of the
two most interesting actions of CRG theory in Sec.~\ref{sec2} --- one
supposedly power-counting renormalizable and the other power-counting
super-renormalizable. Since the actions turn out to contain time
derivatives up to fourth order, we consider ways to simplify the actions
via introduction of additional scalar fields and partial gauge fixing.
In Sec.~\ref{sec3} we analyze the actions using Hamiltonian formalism.
Conclusions and some further discussions are given in Sec.~\ref{sec4}

\section{ADM representation of covariant renormalizable gravity}
\label{sec2}

\subsection{Action}
Let us review the action of the CRG theory proposed in
\cite{Nojiri:2010a,Nojiri:2010b,Nojiri:2010c}. In 4-dimensional
spacetime  the action of the supposedly power-counting renormalizable
gravity is (corresponds to $z=3$)
\be
\label{Sgz3.wT}
S_{g,3} = \int d^4 x \sqrt{-g} \left\{ \frac{R}{2\kappa^2}
- \alpha \left( T^{\mu\nu}R_{\mu\nu} + \beta T R \right)
\left(T^{\mu\nu}\nabla_\mu \nabla_\nu + \gamma T \nabla^\rho
\nabla_\rho\right)
\left( T^{\mu\nu} R_{\mu\nu} + \beta T R \right)
\right\} \,.
\ee
We also consider the supposedly power-counting super-renormalizable
gravity with the action (corresponds to $z=4$)
\be
\label{Sgz4.wT}
S_{g,4} = \int d^4 x \sqrt{-g} \left\{ \frac{R}{2\kappa^2}
- \alpha \left[
\left(T^{\mu\nu}\nabla_\mu \nabla_\nu + \gamma T \nabla^\rho
\nabla_\rho\right)
\left( T^{\mu\nu} R_{\mu\nu} + \beta T R \right) \right]^2
\right\} \,.
\ee
Here $T_{\mu\nu}$ is the energy-momentum tensor of a perfect
non-relativistic fluid with an equation of state parameter $w\neq
1/3,-1$. We denote $T=g^{\mu\nu}T_{\mu\nu}$.
The actions for higher $z$ could be analyzed similarly as these two
interesting representatives of the odd and even $z$ actions.

In Ref.~\cite{Nojiri:2010b} it was proposed that the required perfect
fluid can be realized with a constrained scalar field $\phi$ with the
action
\be
\label{Sphi}
S_\phi = \int d^4 x \sqrt{-g} \left\{ - \lambda \left( \frac{1}{2}
\p_\mu \phi \p^\mu \phi + U(\phi) \right) \right\} \,,
\ee
where $\lambda$ is a Lagrange multiplier field. The energy-momentum
tensor $T_{\mu\nu}$ is defined to be that of an isolated scalar field in
a potential $V(\phi)$,
\be
\label{LagHL4}
T^\phi_{\mu\nu} = \p_\mu \phi \p_\nu \phi
- g_{\mu\nu} \left( \frac{1}{2} \p_\rho \phi \p^\rho \phi
+ V(\phi) \right) \,,
\ee
regardless of the form of the CRG action.
The constraint implied by the action \eqref{Sphi},
\be \label{LMconstraint}
\frac{1}{2}\p_\mu \phi \p^\mu \phi + U(\phi) = 0 \,,
\ee
will be imposed from the start. Thus the constraint
\eqref{LMconstraint}, and consequently the action \eqref{Sphi}, vanish
in the total action of CRG.
According to the constraint \eqref{LMconstraint}, if we assume
$U(\phi)>0$, the vector $\p_\mu \phi$ is timelike. Then at least locally
one can choose the direction of time to be parallel to $\p_\mu \phi$, so
that the constraint \eqref{LMconstraint} yields
\be
\frac{1}{2}\left(\frac{d\phi}{dt}\right)^2 = U(\phi) \,.
\ee
Then on the flat background metric one obtains the energy density
$\rho_\phi=U(\phi)+V(\phi)$, the pressure $p_\phi=U(\phi)-V(\phi)$ and
the equation of state parameter $w=p_\phi/\rho_\phi$ associated with
$T^\phi_{\mu\nu}$.
For simplicity, it is assumed that $V(\phi)$ and $U(\phi)$ are
constants:
\be
\label{LagHL6}
U(\phi) = U_0\,,\qquad V(\phi) = V_0 \,.
\ee

Now the action \eqref{Sgz3.wT} can be written as
\begin{multline} \label{Sgz3}
S_{g,3} = \int d^4 x \sqrt{-g} \left\{ \frac{R}{2\kappa^2} - \alpha
\left(
\p^\mu \phi \p^\nu \phi R_{\mu\nu} + U_0 R \right)
\left(\p^\mu \phi \p^\nu \phi \nabla_\mu \nabla_\nu
+ 2 U_0 \nabla^\rho \nabla_\rho \right) \right. \\
\left. \times \left( \p^\mu \phi \p^\nu \phi R_{\mu\nu} + U_0 R \right)
\right\} \,.
\end{multline}
Likewise, the action for \eqref{Sgz4.wT} is written as
\be \label{Sgz4}
S_{g,4} = \int d^4 x \sqrt{-g} \left\{ \frac{R}{2\kappa^2} - \alpha
\left[
\left(\p^\mu \phi \p^\nu \phi \nabla_\mu \nabla_\nu
+ 2 U_0 \nabla^\rho \nabla_\rho \right) \left(
\p^\mu \phi \p^\nu \phi R_{\mu\nu} + U_0 R \right) \right]^2 \right\}
\,.
\ee
Here the constraint \eqref{LMconstraint} has been used and the following
parameters have been determined:
\be
\beta = -\frac{w-1}{2(3w-1)} = \frac{V_0}{2U_0 - 4V_0} \,,\qquad \gamma
= \frac{1}{3w-1} = \frac{U_0 - V_0}{2U_0 - 4V_0} \,.
\ee
In order to eliminate $U_0$ from the actions \eqref{Sgz3} and
\eqref{Sgz4} we can use the constraint \eqref{LMconstraint} to write
\cite{Nojiri:2010c}
\begin{align}
\p^\mu \phi \p^\nu \phi R_{\mu\nu} + U_0 R &= \p^\mu \phi \p^\nu \phi
\left( R_{\mu\nu} - \frac{1}{2}g_{\mu\nu}R \right) \,,\\
\p^\mu \phi \p^\nu \phi \nabla_\mu \nabla_\nu + 2 U_0 \nabla^\rho
\nabla_\rho &= \p^\mu \phi \p^\nu \phi \left( \nabla_\mu \nabla_\nu -
g_{\mu\nu} \nabla^\rho \nabla_\rho \right) \,.
\end{align}
Since \eqref{Sphi} vanishes, the total action $(S_z=S_{g,z}+S_\phi)$ for
\eqref{Sgz3} can be written
\begin{multline} \label{Sz3}
S_3 = \int d^4 x \sqrt{-g} \left\{ \frac{R}{2\kappa^2} - \alpha \p^\mu
\phi \p^\nu \phi \left( R_{\mu\nu} - \frac{1}{2}g_{\mu\nu}R \right)
\right. \\
\left. \times \p^\mu \phi \p^\nu \phi \left( \nabla_\mu \nabla_\nu -
g_{\mu\nu} \nabla^\rho \nabla_\rho \right) \p^\sigma \phi \p^\lambda
\phi \left( R_{\sigma\lambda} - \frac{1}{2}g_{\sigma\lambda}R \right)
\right\} \,.
\end{multline}
For \eqref{Sgz4} the total action can be written
\begin{multline} \label{Sz4}
S_4 = \int d^4 x \sqrt{-g} \left\{ \frac{R}{2\kappa^2} - \alpha \left[
\p^\mu \phi \p^\nu \phi \left( \nabla_\mu \nabla_\nu - g_{\mu\nu}
\nabla^\rho \nabla_\rho \right) \p^\sigma \phi \p^\lambda \phi \left(
R_{\sigma\lambda} - \frac{1}{2}g_{\sigma\lambda}R \right) \right]^2
\right\} \,.
\end{multline}
The corresponding Lagrangians are denoted by $L_z$, such that $S_z=\int
dt L_z$.

\subsection{Space-time decomposition}
We consider the ADM decomposition of the gravitational field
\cite{Arnowitt:1962} (for reviews and mathematical background, see
\cite{Wald,Gourgoulhon}). Assume that the spacetime admits a foliation
into $t=\mr{constant}$ hypersurfaces $\Sigma_t$, whose future-directed
unit normal is denoted by $n$. The metric tensor $\gM_{ab}$ of spacetime
is decomposed with respect to $\Sigma_t$:
\begin{align}
&g_{ab}=\gM_{ab}-\epsilon n_a n_b \,,& &n_a n^a=\epsilon \,,& &t^a
\nabla_a t=1 \,,\nn\\
&N=\epsilon n_a t^a \,,& &N^a=\projector{a}{b} t^b \,,& &t^a=N n^a+N^a
\,,
\end{align}
where $\epsilon=-1$ ($\epsilon=1$) for a spacelike (timelike)
$\Sigma_t$, and the orthogonal projector from $T\mc{M}$ to $T\Sigma_t$
is
\be \label{projector}
\projector{a}{b} = \delta^a_b - \epsilon n^a n_b \,.
\ee
The extension of tensors on $\Sigma_t$ to tensors on $\mc{M}$ is induced
by the embedding of $\Sigma_t$ to $\mc{M}$, and by the projector map
\eqref{projector}. Such extended quantities are, for example, the metric
tensor $g_{ab}$ of $\Sigma_t$ and the shift vector $N^a$. From now on
quantities defined on the spacetime $\mc{M}$ and associated with its
metric $\gM_{ab}$ are marked with the prefix ${}^{(4)}$. We denote the
covariant derivatives on $\mc{M}$ and $\Sigma_t$ by $\nabla$ and $D$,
respectively.

Taking the scalar field $t$ as the time coordinate, when $t^a$ is the 
``flow of time'' vector, and assuming $\Sigma_t$ is spacelike
($\epsilon=-1$), we can write the unit normal in terms of the lapse $N$
and the shift vector $N^i$:
\be \label{normal}
n_\mu = -N\nabla_\mu t = (-N,0,0,0) \,,\qquad n^\mu = \left(n^0,
n^i\right) = \left(\frac{1}{N}, -\frac{N^i}{N}\right) \,.
\ee
The components of the metric of spacetime are decomposed
\be
\gM_{00} = -N^2+N_i N^i \,,\qquad \gM_{0i} = \gM_{i0} = N_i \,,\qquad
\gM_{ij} = g_{ij}
\ee
and the components of its inverse are
\be
\gM^{00} = -1/N^2 \,,\qquad \gM^{0i} = \gM^{i0} = N^i/N^2 \,,\qquad
\gM^{ij} = g^{ij}-(N^i N^j/N^2) \,.
\ee
The natural volume element decomposes
\be
d^4 x \sqrt{-\gM} = dt d^3\bx \sqrt{g}N \,.
\ee
We assume $c=1$ units.

A 4-vector $u^\mu$ can be decomposed into components perpendicular and
parallel to $n^\mu$:
\be
u^\mu = \per{u}^\mu + \epsilon (n\cdot u) n^\mu \,,\qquad \per{u}^\mu
\equiv \projector{\mu}{\nu} u^\nu = u^\mu-\epsilon (n\cdot u) n^\mu \,.
\ee
For $u^\mu = \p^\mu \phi = \nabla^\mu \phi$ by using \eqref{normal} we
get
\be \label{per_nabla}
\per{u}^\mu = \per{\nabla}^\mu \phi = \left( 0, D^i \phi \right)
\,,\qquad D^i \phi \equiv g^{ij} D_j \phi = g^{ij} \p_j \phi \,,
\ee
and
\be
n\cdot u = \nabla_n \phi \equiv n^\mu \nabla_\mu \phi \,.
\ee
Note that for a vector $\per{u}^\mu \in T\Sigma_t$ and a covector
$\per{v}_\mu \in T^*\Sigma_t$ we have the zero-components
\be
\per{u}^0= 0 \,,\qquad \per{v}_0 = N^i \per{v}_i \,.
\ee
This generalizes straightforwardly for tensors of any type defined on
$\Sigma_t$.

We will apply the ADM decomposition to the actions \eqref{Sz3} and
\eqref{Sz4}.
The scalar curvature $\RM$ of spacetime decomposes as usual:
\begin{align}
\RM &= K_{ij}K^{ij} - K^2 + R + 2\nabla_\mu\left(n^\mu \nabla_\nu n^\nu
- n^\nu \nabla_\nu n^\mu\right) \nn\\
&= K_{ij}K^{ij} - K^2 + R + 2\nabla_\mu\left(n^\mu K\right) -
\frac{2}{N}D^i D_i N \,.\label{RM}
\end{align}
Here $R$ is the intrinsic scalar curvature of $\Sigma_t$ and $K_{ij}$ is
the extrinsic curvature of $\Sigma_t$:
\be\label{K}
K_{ij} = \frac{1}{2N}\left( \dot{g}_{ij} - 2D_{(i}N_{j)} \right)
\,,\qquad K=g^{ij}K_{ij} \,,
\ee
where the dot denotes the derivative with respect to time $t$.

The covariant derivative of the normal covector $n_\mu$ can be written
\be\label{nablan}
\nabla_\mu n_\nu = K_{\mu\nu} - n_\mu D_\nu \ln N \,,
\ee
and it implies the following relations:
\be\label{nablan_implies}
\nabla_n n_\mu = D_\mu \ln N \,,\qquad n^\nu \nabla_\mu n_\nu = 0
\,,\qquad \nabla_\mu n^\mu = K \,.
\ee
These relations have been used in \eqref{RM}. Note that we can also
write
\be \label{nabla_nK}
\nabla_\mu \left(n^\mu K\right)=K^2+\nabla_n K \,,
\ee
for example, in \eqref{RM}.

The Ricci tensor $\RM_{\mu\nu}$ of spacetime can be decomposed as
\be
\RM_{\mu\nu} = \perRM_{\mu\nu} - \perRM_{\mu n}n_\nu -
\perRM_{n\nu}n_\mu + \RM_{nn}n_\mu n_\nu \,,
\ee
where we have defined
\begin{align}
\perRM_{\mu\nu} &\equiv
\projector{\rho}{\mu}\projector{\sigma}{\nu}\RM_{\rho\sigma} =
R_{\mu\nu} + K K_{\mu\nu} - 2K_{\mu\rho}K^\rho_{\phantom{\rho}\nu} -
\frac{1}{N}D_\mu D_\nu N + \frac{1}{N}\mc{L}_{Nn}K_{\mu\nu}
\,,\label{perRmunu}\\
\perRM_{\mu n} &\equiv \projector{\rho}{\mu} n^\nu \RM_{\rho\nu} =
D_\rho K^\rho_{\phantom{\rho}\mu} - D_\mu K \,,\label{Rmun}\\
\perRM_{n\nu} &\equiv \projector{\rho}{\nu} n^\mu \RM_{\mu\rho} = D_\rho
K^\rho_{\phantom{\rho}\nu} - D_\nu K \,,\label{Rnnu}\\
\RM_{nn} &\equiv \RM_{\mu\nu} n^\mu n^\nu = \frac{1}{2} \left( K^2 -
K_{ij}K^{ij} + R - \RM \right) \nn\\
&= K^2 - K_{ij}K^{ij} - \nabla_\mu\left(n^\mu K\right) + \frac{1}{N}D^i
D_i N \,.\label{Rnn}
\end{align}
In Eq. \eqref{perRmunu} $R_{\mu\nu}$ is the Ricci tensor of the
hypersurface $\Sigma_t$ and $\mc{L}_{Nn}$ denotes the Lie derivative
along $Nn^\mu$. Note that for any tensor field $T$ that is tangent to
$\Sigma_t$, $\mc{L}_{Nn}T$ is also tangent to $\Sigma_t$.
In Eqs. \eqref{perRmunu}--\eqref{Rnn} we have used the \emph{Gauss
relation}, the \emph{Ricci equation} and the \emph{Codazzi relation}, 
and in \eqref{Rnn} the decomposition of $\RM$ from \eqref{RM} was also
used. Hence the Einstein tensor can be decomposed:
\begin{multline}
\GM_{\mu\nu} \equiv \RM_{\mu\nu} - \frac{1}{2}\gM_{\mu\nu}\RM =
R_{\mu\nu} + K K_{\mu\nu} - 2K_{\mu\rho}K^\rho_{\phantom{\rho}\nu} -
\frac{1}{N}D_\mu D_\nu N + \frac{1}{N}\mc{L}_{Nn}K_{\mu\nu} \\
- \frac{1}{2}g_{\mu\nu} \left( R + K_{ij}K^{ij} + K^2 + 2\nabla_n K -
\frac{2}{N}D^i D_i N \right) \\
+ \left( D_\mu K - D_\rho K^\rho_{\phantom{\rho}\mu} \right) n_\nu
+ n_\mu \left( D_\nu K - D_\rho K^\rho_{\phantom{\rho}\nu} \right)
+ \frac{1}{2} n_\mu n_\nu \left( K^2 - K_{ij}K^{ij} + R \right) \,,
\end{multline}
where we have also used \eqref{nabla_nK}.
Thus in the actions \eqref{Sz3} and \eqref{Sz4} we have
\begin{multline}\label{ppG}
\p^\mu \phi \p^\nu \phi \, \GM_{\mu\nu} = D^i \phi D^j \phi \left[
R_{ij} + K K_{ij} - 2K_{ik}K^k_{\phantom{k}j} - \frac{1}{N}D_i D_j N +
\frac{1}{N}\mc{L}_{Nn}K_{ij} \right.\\
-\left. \frac{1}{2} g_{ij} \left( R + K_{kl}K^{kl} + K^2 + 2\nabla_n K -
\frac{2}{N}D^k D_k N \right) \right] \\
+ 2\left(\nabla_n \phi\right) D^i \phi \left( D_i K - D^j K_{ji} \right)
+ \frac{1}{2} \left(\nabla_n \phi\right)^2 \left( K^2 - K_{ij}K^{ij} + R
\right) \,.
\end{multline}

Let us then consider the covariant derivatives in the actions
\eqref{Sz3} and \eqref{Sz4}. The covariant derivative $D$ of a
$(k,l)$-tensor field $T$ on $\Sigma_t$ is given by
\be \label{DmuT}
D_\mu
T^{\nu_1\cdots\nu_k}_{\phantom{\nu_1\cdots\nu_k}\rho_1\cdots\rho_l} =
\projector{\sigma}{\mu}
\projector{\nu_1}{\alpha_1}\cdots\projector{\nu_k}{\alpha_k}
\projector{\beta_1}{\rho_1}\cdots\projector{\beta_l}{\rho_l}
\nabla_\sigma
T^{\alpha_1\cdots\alpha_k}_{\phantom{\alpha_1\cdots\alpha_k}
\beta_1\cdots\beta_l} \,,
\ee
where in the right-hand side one considers the extension of $T$ on
$\mc{M}$. For a scalar field $f$, $D_\mu
f=\projector{\nu}{\mu}\nabla_\nu f$ implies the decomposition
\be
\nabla_\mu f = D_\mu f - n_\mu \nabla_n f \,.
\ee
For the second order covariant derivative we obtain
\be \label{DDf}
D_\mu D_\nu f = \projector{\rho}{\mu}\projector{\sigma}{\nu} \nabla_\rho
\nabla_\sigma f + \projector{\rho}{(\mu}\projector{\sigma}{\nu)}
\left(\nabla_\rho \projector{\lambda}{\sigma}\right) \nabla_\lambda f
\,.
\ee
Note that both $D$ and $\nabla$ are torsion-free, which is the reason
why the last term is symmetrized over $\mu$ and $\nu$. Introducing
\eqref{projector} into \eqref{DDf} yields
\begin{multline}
\nabla_\mu \nabla_\nu f = D_\mu D_\nu f - 2n_{(\mu|} \nabla_n
\nabla_{|\nu)} f - n_\mu n_\nu n^\rho n^\sigma \nabla_\rho \nabla_\sigma
f \\
- \left( \nabla_{(\mu} n_{\nu)} + n_{(\mu|}\nabla_n n_{|\nu)} +
n_{(\mu}n^\rho \nabla_{\nu)}n_\rho + n_\mu n_\nu n^\rho \nabla_n n_\rho
\right) \nabla_n f \,.
\end{multline}
Then we shall use the relations \eqref{nablan} and
\eqref{nablan_implies} to obtain
\be
\nabla_\mu \nabla_\nu f = D_\mu D_\nu f - K_{\mu\nu}\nabla_n f -
2n_{(\mu}\projector{\rho}{\nu)} \nabla_n \nabla_\rho f + n_\mu n_\nu
n^\rho n^\sigma \nabla_\rho \nabla_\sigma f \,.
\ee
As the last step in the decomposition we write
\be
\nabla_n \nabla_\rho f = \nabla_\rho \nabla_n f - \nabla_\rho n^\sigma
\nabla_\sigma f = \nabla_\rho \nabla_n f -
\left(K_\rho^{\phantom{\rho}\sigma}-n_\rho D^\sigma \ln N\right)
\nabla_\sigma f \,,
\ee
which implies
\begin{align}
\projector{\rho}{\nu} \nabla_n \nabla_\rho f &= D_\nu \nabla_n f -
K_\nu^{\phantom{\nu}\rho} \nabla_\rho f \,,\\
n^\rho n^\sigma \nabla_\rho \nabla_\sigma f &= \nabla_n \nabla_n f -
(D^i \ln N) D_i f \,.
\end{align}
Thus the final decomposition is
\begin{multline}
\label{nabla2f}
\nabla_\mu \nabla_\nu f = D_\mu D_\nu f - K_{\mu\nu}\nabla_n f -
2n_{(\mu} D_{\nu)} \nabla_n f - 2n_{(\mu}K_{\nu)}^{\phantom{\nu)}\rho}
\nabla_\rho f \\
+ n_\mu n_\nu \left( \nabla_n \nabla_n f - (D^i \ln N) D_i f \right) \,.
\end{multline}
The contraction of \eqref{nabla2f} with $\gM^{\mu\nu}$ gives
\be
\nabla^\mu \nabla_\mu f = D^i D_i f - K \nabla_n f - \nabla_n \nabla_n f
+ (D^i \ln N) D_i f \,,
\ee
where we have also used
\be
\gM^{\mu\nu} D_\mu D_\nu f = g^{ij}D_i D_j f = D^i D_i f \,,\qquad
\gM^{\mu\nu} K_{\mu\nu} = g^{ij} K_{ij} = K \,.
\ee
Then in the action we have
\begin{multline}\label{ppnabla2}
\p^\mu \phi \p^\nu \phi \left( \nabla_\mu \nabla_\nu - \gM_{\mu\nu}
\nabla^\rho \nabla_\rho \right) f = D^i \phi D^j \phi \left[ \left( D_i
D_j - g_{ij}D^k D_k - g_{ij}(D^k \ln N) D_k \right) f \right. \\
\left. - \left( K_{ij} - g_{ij}K \right) \nabla_n f + g_{ij} \nabla_n
\nabla_n f \right] \\
- 2\left(\nabla_n \phi\right) D^i \phi \left( D_i \nabla_n + K_{ij}D^j
\right) f
+ \left(\nabla_n \phi\right)^2 \left( D^i D_i - K \nabla_n \right) f \,.
\end{multline}

The covariant derivative of a scalar $f$ along the normal $n$ is given
by
\be \label{nabla_n_f}
\nabla_n f = \frac{1}{N}\left( \dot{f} - N^i \p_i f \right) \,.
\ee

Now we have everything that is needed for the ADM decomposition of the
actions of CRG. First, in order to shorten the expressions, we denote
the decomposition \eqref{ppG} by
\be
\GMphi \equiv \p^\mu \phi \p^\nu \phi \GM_{\mu\nu} \,.
\ee
Then the action \eqref{Sz3} is written in terms of the ADM variables:
\begin{multline} \label{Sz3.ADM}
S_3 = \int dt d^3\bx \sqrt{g}N \left\{
\frac{K_{ij}K^{ij}-K^2+R}{2\kappa^2} - \alpha \GMphi \left[
D^i \phi D^j \phi \left[ \left( D_i D_j - g_{ij}D^k D_k
\right.\right.\right.\right. \\
\left.\left. - g_{ij}(D^k \ln N) D_k \right) \GMphi - \left( K_{ij} -
g_{ij}K \right) \nabla_n \GMphi + g_{ij} \nabla_n \nabla_n \GMphi
\right] \\
\left.\left. - 2\left(\nabla_n \phi\right) D^i \phi \left( D_i \nabla_n
+ K_{ij}D^j \right) \GMphi
+ \left(\nabla_n \phi\right)^2 \left( D^i D_i - K \nabla_n \right)
\GMphi \right] \right\} \,.
\end{multline}
The action \eqref{Sz4} can be written as
\begin{multline} \label{Sz4.ADM}
S_4 = \int dt d^3\bx \sqrt{g}N \left\{
\frac{K_{ij}K^{ij}-K^2+R}{2\kappa^2} - \alpha \left[
D^i \phi D^j \phi \left[ \left( D_i D_j - g_{ij}D^k D_k
\right.\right.\right.\right. \\
\left.\left. - g_{ij}(D^k \ln N) D_k \right) \GMphi - \left( K_{ij} -
g_{ij}K \right) \nabla_n \GMphi + g_{ij} \nabla_n \nabla_n \GMphi
\right] \\
\left.\left. - 2\left(\nabla_n \phi\right) D^i \phi \left( D_i \nabla_n
+ K_{ij}D^j \right) \GMphi
+ \left(\nabla_n \phi\right)^2 \left( D^i D_i - K \nabla_n \right)
\GMphi \right]^2 \right\} \,.
\end{multline}
We assume that the boundary terms originating from total derivatives
vanish when appropriate boundary conditions are imposed, even though the
role of surface integrals is known to be very important in general
relativity.
Introducing the decomposition \eqref{ppG} and the covariant derivative
along $n$ \eqref{nabla_n_f} explicitly into \eqref{Sz3.ADM} would
clearly result into a very complicated expression. The action
\eqref{Sz3.ADM} contains time derivatives of the metric $g_{ij}$ up to
fourth order, time derivatives of the lapse $N$ and the shift vector
$N^i$ (or of its spatial derivative $D_i N_j$)  up to third order, and
time derivatives of $\phi$ up to third order. Some higher time
derivatives of the ADM variables could be removed via integration by
parts, but with the price of introducing even higher time derivatives of
the scalar field $\phi$.

For the Lagrangian $L_3$ defined by the action \eqref{Sz3.ADM} we find
\be
\frac{\delta^2 L_3}{\delta g_{ij}^{(4)}\delta g_{kl}^{(4)}} = 0 \,,
\ee
where $g_{ij}^{(n)} \equiv \frac{\p^n g_{ij}}{\p t^n}$. This is clear
since the highest time derivative is contained in the term
\be
\alpha \GMphi (D^i \phi D_i \phi)^2 \nabla_n \nabla_n \nabla_n K \,,
\ee
where
\be
\nabla_n \nabla_n \nabla_n K = \frac{g^{ij}}{2N^4} g_{ij}^{(4)} +
\text{(terms with lower order time derivatives)} \,.
\ee
Thus $g_{ij}^{(4)}$ cannot be solved in terms of the canonical
variables, nor $g_{ij}^{(8)}$ defined in terms of the initial data from
the equations of motion.

For the Lagrangian $L_4$ defined by the action \eqref{Sz4.ADM} the
highest time derivative is contained in the term
\be
-\alpha (D^i \phi D_i \phi)^3 (\nabla_n \nabla_n \nabla_n K)^2
\ee
and
\be
\frac{\delta^2 L_4}{\delta g_{ij}^{(4)}\delta g_{kl}^{(4)}} = -
\sqrt{g}N \frac{g^{ij}g^{kl}}{4N^8} \alpha (D^m \phi D_m \phi)^3 \,.
\ee
Therefore we expect that $g_{ij}^{(4)}$ could be solved in terms of the
canonical variables.

However, since neither of the actions \eqref{Sz3.ADM} or \eqref{Sz4.ADM}
depend on $N^{(4)}$, $N^{i(4)}$ or $\phi^{(4)}$, the highest order
Hessian matrices
\be
\frac{\delta^2 L_3}{\delta q_I^{(4)}\delta q_J^{(4)}} \quad\mr{and}\quad
\frac{\delta^2 L_4}{\delta q_I^{(4)}\delta q_J^{(4)}} \,,
\ee
where $q_I=(\phi, N, N^i, g_{ij}), I=1,2,\ldots,11$, have the ranks zero
and six everywhere, respectively. Thus there are eleven and five
Lagrangian constraints associated with $L_3$ and $L_4$, respectively.
As expected, both Lagrangians are singular. The reason for the
singularity is the presence of gauge symmetry, and its associated
(first-class) constraints.

The actions proposed in Refs.~\cite{Nojiri:2010b,Nojiri:2010c} for
higher $z$ ($z>4$) can also be written in the ADM form by using the
decompositions we have obtained above, \eqref{ppG} and \eqref{ppnabla2}
in particular. Then we see that for the Lagrangian of even $z=2n+2$ and
of odd $z=2n+3$ the highest order Hessian matrix is singular, similarly
to the two cases discussed above.

\subsection{Partial gauge fixing} \label{sec2.3}
The constraint \eqref{LMconstraint} on $\phi$ ensures that the vector
$\p^\mu \phi$ is timelike everywhere, when $U(\phi)>0$ is assumed.
Therefore there exists a preferred foliation of space-time into spatial
hypersurfaces $\Sigma_t$ whose unit normal is given by
\be \label{n:pphi}
n^\mu = -\frac{\p^\mu \phi}{\sqrt{-\p_\nu \phi \p^\nu \phi}} =
-\frac{\p^\mu \phi}{\sqrt{2U(\phi)}} \,.
\ee
Hence we can write
\be \label{pphi}
\p_\mu \phi = -\sqrt{2U(\phi)} n_\mu \,,\qquad \p^\mu \phi =
-\sqrt{2U(\phi)} n^\mu \,,
\ee
where the unit normal is given in terms of the ADM variables in Eq.
\eqref{normal}. From Eq. \eqref{pphi} we see that in this foliation
$\phi$ is constant on $\Sigma_t$, $\phi=\phi(t)$.
Thus the constraint \eqref{LMconstraint} on $\phi$ reduces to
\be \label{LMconstraint.gf}
-\frac{\dot{\phi}^2}{2N^2} + U(\phi) = 0 \,.
\ee
This implies that the lapse $N$ must be constant on $\Sigma_t$ too,
$N=N(t)$. In order to preserve this condition we restrict the symmetry
under diffeomorphisms of space-time to the symmetry under
foliation-preserving diffeomorphisms, given in the infinitesimal form as
\be \label{fp-diffeomorphism}
\delta t = f(t) \,,\qquad \delta\bx = \bm{\xi}(t,\bx) \,.
\ee
This is the main symmetry group of the HL gravity \cite{Horava:2009uw}. 
In the language of Ho\v{r}ava's theory we would say that both $\phi$ and
$N$ are projectable --- like the lapse $N$ is in projectable HL gravity.
Here we consider \eqref{fp-diffeomorphism} as a partial gauge fixing of
the diffeomorphism symmetry.

Now the Eqs.~\eqref{ppG} and \eqref{ppnabla2} can be written
\begin{gather}
\p^\mu \phi \p^\nu \phi \GM_{\mu\nu} = U_0 \left(K^2 - K_{ij}K^{ij} + R
\right) \,,\label{ppG.gf}\\
\p^\mu \phi \p^\nu \phi \left( \nabla_\mu \nabla_\nu - \gM_{\mu\nu}
\nabla^\rho \nabla_\rho \right) f = 2U_0 \left( D^i D_i - K \nabla_n
\right) f \,,\label{ppnabla2.gf}
\end{gather}
where \eqref{LagHL6} has again been assumed. 
Note that $\phi$ is no longer an independent variable, since the
constraint \eqref{LMconstraint.gf} implies
\be
\phi(t) = \phi(t_0) + \sqrt{2U_0} \int_{t_0}^t dt' N(t') \,.
\ee
Moreover, $\phi$ is not even present in the actions anymore, as we will
see next.
The actions \eqref{Sz3.ADM} and \eqref{Sz4.ADM} reduce to
\begin{multline} \label{Sz3.ADM.gf}
S_3 = \int dt d^3\bx \sqrt{g}N \left\{
\frac{K_{ij}K^{ij}-K^2+R}{2\kappa^2} - \alpha 2U_0^3 \left(K^2 -
K_{ij}K^{ij} + R \right) \right. \\
\left. \times \left( D^i D_i - K \nabla_n \right) \left(K^2 -
K_{ij}K^{ij} + R \right) \right\} \,,
\end{multline}
and
\begin{multline} \label{Sz4.ADM.gf}
S_4 = \int dt d^3\bx \sqrt{g}N \left\{
\frac{K_{ij}K^{ij}-K^2+R}{2\kappa^2} \right. \\
\left. - \alpha \left[ 2U_0^2 \left( D^i D_i - K \nabla_n \right)
\left(K^2 - K_{ij}K^{ij} + R \right) \right]^2 \right\} \,.
\end{multline}
Although these actions have been simplified considerably, they still
contain second order time derivatives of the metric $g_{ij}$, and first
order time derivatives of the lapse $N$ and the shift vector $N^i$.
Indeed, we obtain
\begin{multline}
\nabla_n \left(K^2 - K_{ij}K^{ij} + R \right) = \frac{1}{N} \left[
2\left(\dot{K}- N^i \p_i K\right) K - 2\left(\dot{K}_{ij} - N^k \p_k
K_{ij}\right) K^{ij} \right. \\
\left. - 2\left(\dot{g}^{ij}- N^k \p_k g^{ij}\right)
K_{il}K_j^{\phantom{j}l} + \left( \dot{R} - N^i \p_i R \right) \right]
\,,
\end{multline}
where the time derivatives of the extrinsic curvature \eqref{K} and the
inverse metric $g^{ij}$ are given by
\be
\dot{K}_{ij} = - \frac{\dot{N}}{N} K_{ij} + \frac{1}{2N}\left(
\ddot{g}_{ij} - 2\p_t D_{(i}N_{j)} \right) \,,
\ee
\be
\dot{K} = \dot{g}^{ij}K_{ij} + g^{ij}\dot{K}_{ij} \,,\qquad \dot{g}^{ij}
= - g^{ik}g^{jl}\dot{g}_{kl} \,.
\ee

We can see that our earlier observations on the singularity of $L_3$ and
$L_4$ have not changed. For the Lagrangian $L_3$ we obtain
\be
\frac{\delta^2 L_3}{\delta\ddot{g}_{ij}\delta\ddot{g}_{kl}} = 0 \,.
\ee
For the Lagrangian $L_4$ we obtain
\be
\frac{\delta^2 L_4}{\delta\ddot{g}_{ij}\delta\ddot{g}_{kl}} = -
\sqrt{g}\alpha\frac{8U_0^4}{N^3} K^2 \left(K^{ij}-g^{ij}K\right)
\left(K^{kl}-g^{kl}K\right)\,.
\ee
Thus for the Hessian matrices we find
\be
\rank \left[ \frac{\delta^2 L_3}{\delta \ddot{q}_I \delta \ddot{q}_J}
\right] = 0 \,,\qquad
\rank \left[ \frac{\delta^2 L_4}{\delta \ddot{q}_I \delta \ddot{q}_J}
\right] = 6 \,,
\ee
and hence we expect ten and four Lagrangian constraints for $L_3$ and
$L_4$, respectively.

Second order time derivatives in the Lagrangian of a theory of gravity
are not necessarily a fatal problem. Indeed we know that actions of the
type $f(\RM)$ do not suffer from the Ostrogradskian instability (for a
proper explanation, see the discussion in Ref.~\cite{Woodard:2007}),
though the scalar curvature \eqref{RM} contains the second order time
derivative term
\be
2\nabla_\mu\left(n^\mu K\right) = \frac{g^{ij}\ddot{g}_{ij}}{N^2} +
\text{(terms with first order time derivatives)}\,.
\ee
In the $f(\RM)$ action one can get rid of the second order time
derivative by introducing additional scalar fields and integrating by
parts, although with the price of an extra dynamical scalar degree of
freedom --- the scalaron. In the present case a somewhat similar
approach can reduce the number of time derivatives in the action. This
will be discussed next.

\subsection{Further reducing the number of time derivatives}
\label{sec2.4}
As discussed in the Appendix of Ref.~\cite{Nojiri:2010c}, in order to
reduce the number of derivatives in the action, one can introduce four
scalar fields $\zeta_1$, $\xi_1$, $\zeta_2$ and $\xi_2$, and write the
action \eqref{Sz3} in the form \cite{Nojiri:2010c}:
\begin{multline} \label{Sz3.st}
S_3 = \int d^4 x \sqrt{-\gM}\left\{ \frac{\RM}{2\kappa^2} - \alpha
\left[ \zeta_1 \zeta_2 + \xi_1 \left( \zeta_1 - \p^\mu \phi \p^\nu \phi
\left( \nabla_\mu \nabla_\nu - \gM_{\mu\nu} \nabla^\rho \nabla_\rho
\right) \zeta_2 \right) \right.\right. \\
+ \left. \left. \xi_2 \left( \zeta_2 - \p^\mu \phi \p^\nu \phi
\GM_{\mu\nu} \right) \right] \right\} \,.
\end{multline}
For the action \eqref{Sz4} we introduce three scalar field $\eta, \zeta,
\xi$ in order to write:
\begin{multline} \label{Sz4.st}
S_4 = \int d^4 x \sqrt{-\gM}\left\{ \frac{\RM}{2\kappa^2} + \alpha
\left[ \eta^2 - 2\eta \p^\mu \phi \p^\nu \phi \left( \nabla_\mu
\nabla_\nu - \gM_{\mu\nu} \nabla^\rho \nabla_\rho \right) \zeta
\right.\right. \\
+ \left. \left. \xi \left( \zeta - \p^\mu \phi \p^\nu \phi \GM_{\mu\nu}
\right) \right] \right\} \,.
\end{multline}
The bad news is that the actions \eqref{Sz3.st} and \eqref{Sz4.st} still
contain second order time derivatives of the scalar fields $\zeta_2$ and
$\zeta$, respectively.

However, if we employ the preferred foliation discussed in
Sec.~\ref{sec2.3}, we obtain actions with only first order derivatives
with respect to time. Moreover, the kinetic terms of the actions are
quadratic in the first time derivatives of the fields. Indeed, we obtain
the following ADM representations of the actions \eqref{Sz3.st} and
\eqref{Sz4.st}:
\begin{multline} \label{Sz3.ADM.st.gf}
S_3 = \int dt d^3\bx \sqrt{g}N \left\{
\frac{\mc{G}^{ijkl}K_{ij}K_{kl}+R}{2\kappa^2} - \alpha  \left[ \zeta_1
\zeta_2 + \xi_1 \left( \zeta_1 + 2U_0 \left( K \nabla_n - D^i D_i
\right) \zeta_2 \right) \right.\right. \\
+ \left.\left. \xi_2 \left( \zeta_2 + U_0 \left(
\mc{G}^{ijkl}K_{ij}K_{kl}-R \right) \right) \right] \right\}
\end{multline}
and
\begin{multline} \label{Sz4.ADM.st.gf}
S_4 = \int dt d^3\bx \sqrt{g}N \left\{
\frac{\mc{G}^{ijkl}K_{ij}K_{kl}+R}{2\kappa^2} + \alpha \left[ \eta^2 +
4U_0 \eta \left( K \nabla_n - D^i D_i \right) \zeta \right.\right. \\
+ \left.\left. \xi \left( \zeta + U_0 \left( \mc{G}^{ijkl}K_{ij}K_{kl}-R
\right) \right) \right] \right\} \,.
\end{multline}
Here we have introduced the so-called De Witt metric
\be
\mc{G}^{ijkl} = \frac{1}{2}\left( g^{ik}g^{jl}+g^{il}g^{jk} \right) -
g^{ij}g^{kl} \,,
\ee
which is the special case $\sigma=1$ of the ``generalized De Witt
metric''
\be \label{genDeWitt}
\mc{G}_{(\sigma)}^{ijkl} = \frac{1}{2}\left( g^{ik}g^{jl}+g^{il}g^{jk}
\right) - \sigma g^{ij}g^{kl} \,.
\ee
When $\sigma\neq1/3$, \eqref{genDeWitt} has the inverse
\be
\mc{G}_{(\sigma)ijkl} = \frac{1}{2}\left( g_{ik}g_{jl}+g_{il}g_{jk}
\right) - \frac{\sigma}{3\sigma-1} g_{ij}g_{kl} \,,\qquad
\mc{G}_{(\sigma)ijkl}\mc{G}_{(\sigma)}^{klmn} =
\delta_{(i}^{(m}\delta_{j)}^{n)} \,.
\ee
Note that the actions \eqref{Sz3.ADM.st.gf} and \eqref{Sz4.ADM.st.gf} no
longer contain time derivatives of the lapse $N$ and the shift vector
$N^i$. This means that the lapse and the shift vector are nondynamical,
and hence do not propagate, which is also an attractive feature.
Obviously the Lagrangians of \eqref{Sz3.ADM.st.gf} and
\eqref{Sz4.ADM.st.gf} are still singular due to the remaining gauge
symmetry \eqref{fp-diffeomorphism}. The Hessian matrix for the
Lagrangian $L_3$ of \eqref{Sz3.ADM.st.gf} is given by
\be
\frac{\delta^2 L_3}{\delta\dot{g}_{ij}\delta\dot{g}_{kl}} = \sqrt{g}
\left( \frac{1}{2\kappa^2}-\alpha U_0 \xi_2 \right) \frac{1}{2N}
\mc{G}^{ijkl} \,,\qquad
\frac{\delta^2 L_3}{\delta\dot{\zeta}_2\delta\dot{g}_{ij}} = -
\sqrt{g}\alpha U_0 \frac{g^{ij}\xi_1}{N} \,,
\ee
and the rest of the components are zero.
The nonvanishing components of the Hessian matrix for the Lagrangian
$L_4$ of \eqref{Sz4.ADM.st.gf} are
\be
\frac{\delta^2 L_4}{\delta\dot{g}_{ij}\delta\dot{g}_{kl}} = \sqrt{g}
\left( \frac{1}{2\kappa^2}+\alpha U_0 \xi \right) \frac{1}{2N}
\mc{G}^{ijkl} \,,\qquad
\frac{\delta^2 L_4}{\delta\dot{\zeta}\delta\dot{g}_{ij}} =
\sqrt{g}\alpha 2U_0 \frac{g^{ij}\eta}{N} \,.
\ee
Thus we expect to have seven and six Lagrangian constraints for $L_3$
and $L_4$, respectively.

\section{Hamiltonian formalism}
\label{sec3}
A Hamiltonian formalism for higher derivative theories was first
developed by Ostrogradski \cite{Ostrogradski:1850}. Regular Lagrangians
with higher order time derivatives are known to possess negative energy
degrees of freedom that destabilize solutions of the theory
\cite{Ostrogradski:1850,Eliezer:1989}. But gauge theories are never
regular, so Dirac's formalism \cite{Dirac:1950,Dirac:1958,Dirac:1964},
which is suited for constrained (singular) systems, has to be used
instead. Generalization of Dirac's formalism to higher derivative
theories, and a generalization of Ostrogradski's formalism to singular
systems, were finally achieved in Ref.~\cite{Pons:1989}.

Lagrangians with higher order time derivatives have also been studied in
the context of nonlocal theories. In particular see \cite{Llosa:1994},
though only regular higher order Lagrangians are considered there,
whereas all the CRG Lagrangians are singular. Theories where higher
derivative terms are regarded as corrections to a corresponding lower
order theory, and especially theories that are truncated perturbative
expansions of a nonlocal theory, should employ so-called perturbative
constraints that ensure that the higher order corrections do not change
the theory qualitatively (e.g. introduce extra degrees of freedom, lack
of lower energy bound or violation of unitarity)
\cite{Eliezer:1989,Simon:1990}.

GR with a minimal coupling to a scalar $\phi$ that is constrained by the
same Lagrange multiplier constraint \eqref{LMconstraint} as in CRG has
been proposed in Ref.~\cite{Capozziello:2010}. Its Hamiltonian formalism
has been studied in Ref.~\cite{Kluson:2010}.

The Hamiltonian analysis of the actions \eqref{Sz3.ADM} and
\eqref{Sz4.ADM}, which contain time derivatives up to fourth order,
could be done using the formalism of Ref.~\cite{Pons:1989}. Likewise, we
could analyze the simpler actions \eqref{Sz3.ADM.gf} and
\eqref{Sz4.ADM.gf} or \eqref{Sz3.ADM.st.gf} and \eqref{Sz4.ADM.st.gf},
which contain second order time derivatives. Here we, however,
concentrate on the analysis of the first order actions
\eqref{Sz3.ADM.st.gf} and \eqref{Sz4.ADM.st.gf}. The motivation is to
gain some understanding from these simplified CRG actions, because the
general ones have proven to be quite difficult to analyze.

\subsection{The action \texorpdfstring{$S_3$}{S3}}
First let us define the canonical momenta. Since the action
\eqref{Sz3.ADM.st.gf} is independent of the time derivatives of $N$,
$N^i$, $\zeta_1$, $\xi_1$ and $\xi_2$, their canonically conjugated
momenta, $p_N$, $p_i$, $p_{\zeta_1}$, $p_{\xi_1}$ and $p_{\xi_2}$,
respectively, are the primary constraints:
\be \label{pconstraints_z3}
p_N \approx 0 \,,\qquad p_i(\bx) \approx 0 \,,\qquad p_{\zeta_1}(\bx)
\approx 0 \,,\qquad p_{\xi_1}(\bx) \approx 0 \,,\qquad p_{\xi_2}(\bx)
\approx 0 \,.
\ee
The momenta canonically conjugate to $g_{ij}$ and $\zeta_2$ are defined
by
\begin{align}
p^{ij} &= \frac{\delta S_3}{\delta\dot{g}_{ij}} = \sqrt{g}\left(
\frac{1}{2\kappa^2} \mc{G}^{ijkl}K_{kl} - \alpha U_0 \left( \xi_1 g^{ij}
\nabla_n \zeta_2 + \xi_2 \mc{G}^{ijkl}K_{kl} \right) \right)
\,,\label{p^ij_z3}\\
p_{\zeta_2} &= \frac{\delta S_3}{\delta\dot{\zeta}_2} = - \sqrt{g}
2\alpha U_0 \xi_1 K \,.\label{p_zeta2}
\end{align}
The Poisson bracket (PB) between the fields and the momenta are
postulated in the conventional form. The nonvanishing PBs are defined
(equal time $t$ is understood):
\begin{align}
&\{ g_{ij}(\bx), p^{kl}(\by) \} =
\delta_{(i}^{(k}\delta_{j)}^{l)}\delta^3(\bx-\by) \,,&
&\{ N, p_N \} = 1 \,,\nn\\
&\{ N^i(\bx), p_j(\by) \} = \delta^i_j \delta^3(\bx-\by) \,,&
&\{ \zeta_1(\bx), p_{\zeta_1}(\by) \} = \delta^3(\bx-\by)
\,,\label{PB_z3}
\end{align}
and similarly for the fields $\xi_1$, $\zeta_2$ and $\xi_2$ as for
$\zeta_1$. Recall that the lapse variable was constrained to be
projectable, $N=N(t)$, due to the constraints \eqref{LMconstraint.gf}
and $\phi=\phi(t)$.

In order to obtain the Hamiltonian, we perform the Legendre transform in
the usual manner. First we solve $\dot{\zeta}_2$ and $\dot{g}_{ij}$ in
terms of the canonical variables (the fields and their conjugated
momenta):
\begin{align}
\dot{\zeta}_2 &= N^i \p_i \zeta_2 - \frac{N}{\sqrt{g}3\alpha U_0 \xi_1}
\left( p - \frac{1-2\kappa^2 \alpha U_0 \xi_2}{2\kappa^2 \alpha U_0
\xi_1} p_{\zeta_2} \right) \,,\label{dotzeta_2}\\
\dot{g}_{ij} &= 2NK_{ij} + 2D_{(i}N_{j)} \,,\\
K_{ij} &= \frac{1}{\sqrt{g}} \frac{2\kappa^2}{1-2\kappa^2 \alpha U_0
\xi_2} \mc{G}_{ijkl} \left[ p^{kl}-\frac{g^{kl}}{3}\left( p -
\frac{1-2\kappa^2 \alpha U_0 \xi_2}{2\kappa^2 \alpha U_0 \xi_1}
p_{\zeta_2} \right) \right] \,,\label{K_ij_z3}
\end{align}
where we denote $p=g_{ij}p^{ij}$. Then we define the total Hamiltonian
by
\be \label{H_3.def}
H_3 = \int d^3\bx \left( \dot{g}_{ij}p^{ij} + \dot{\zeta}_2 p_{\zeta_2}
+ \lambda_N p_N + \lambda^i p_i + \lambda_{\zeta_1}p_{\zeta_1} +
\lambda_{\xi_1}p_{\xi_1} + \lambda_{\xi_2}p_{\xi_2} \right) - L_3 \,,
\ee
where $\lambda_N$, $\lambda^i$, $\lambda_{\zeta_1}$, $\lambda_{\xi_1}$
and $\lambda_{\xi_2}$ are Lagrange multiplier fields. We obtain the
Hamiltonian by substituting \eqref{dotzeta_2}--\eqref{K_ij_z3} into
\eqref{H_3.def}:
\be \label{H_3.res}
H_3 = \int d^3\bx \left( N\mc{H}^3_0 + N^i \mc{H}^3_i + \lambda_N p_N +
\lambda^i p_i + \lambda_{\zeta_1}p_{\zeta_1} + \lambda_{\xi_1}p_{\xi_1}
+ \lambda_{\xi_2}p_{\xi_2} \right) \,,
\ee
where we have defined
\begin{align}
\mc{H}^3_0 &= \frac{1}{\sqrt{g}} \frac{2\kappa^2}{1-2\kappa^2 \alpha U_0
\xi_2} \left[ p_{ij}p^{ij} - \frac{1}{3}p^2 - \frac{1}{3}\left(
\frac{1-2\kappa^2 \alpha U_0 \xi_2}{2\kappa^2 \alpha U_0 \xi_1}\right) p
p_{\zeta_2} + \frac{1}{6}\left( \frac{1-2\kappa^2 \alpha U_0
\xi_2}{2\kappa^2 \alpha U_0 \xi_1}\right)^2 p_{\zeta_2}^2 \right] \nn\\
&+ \sqrt{g}\left\{ -\frac{1+2\kappa^2 \alpha U_0 \xi_2}{2\kappa^2} R +
\alpha \left[ \zeta_1 \zeta_2 + \xi_1 \left( \zeta_1-2U_0 D^i D_i
\zeta_2 \right) + \xi_2 \zeta_2 \right] \right\} \,,\\
\mc{H}^3_i &= - 2g_{ij}D_k p^{jk} + D_i \zeta_2 p_{\zeta_2}\nn\\
&= -2g_{ij}\p_k p^{jk} - \left( 2\p_j g_{ik} - \p_i
g_{jk} \right) p^{jk} + \p_i \zeta_2 p_{\zeta_2}
\,.\label{momentum-constraints}
\end{align}
Here we denote $p_{ij}=g_{ik}g_{jl}p^{kl}$. $\mc{H}^3_i$ are the
momentum constraints --- three constraints at every point on $\Sigma_t$.
Because of the condition $N=N(t)$ the Hamiltonian constraint is the
integral of $\mc{H}^3_0$ --- a single global constraint.

The primary constraints \eqref{pconstraints_z3} must be preserved under
time evolution generated by the total Hamiltonian $H_3$ of the system:
\begin{align}
\dot{p}_N &= \{ p_N, H_3 \} = -\int d^3\bx \mc{H}^3_0 \,,\\
\dot{p}_i &= \{ p_i, H_3 \} = -\mc{H}^3_i \,,\\
\dot{p}_{\zeta_1} &= \{ p_{\zeta_1}, H_3 \} = -N \{ \int d^3\bx
\mc{H}^3_0, p_{\zeta_1} \} = -N\sqrt{g}\alpha \left( \zeta_2+\xi_1
\right) \,,\label{dotp_zeta_1}\\
\dot{p}_{\xi_1} &= \{ p_{\xi_1}, H_3 \} = -N \{ \int d^3\bx \mc{H}^3_0,
p_{\xi_1} \} \nn\\
&= -N \left[ \frac{1}{\sqrt{g}}\frac{1}{3\alpha U_0 \xi_1^2} \left(
pp_{\zeta_2} - \frac{1-2\kappa^2 \alpha U_0 \xi_2}{2\kappa^2 \alpha U_0
\xi_1} p_{\zeta_2}^2 \right) + \sqrt{g}\alpha \left( \zeta_1 - 2U_0 D^i
D_i \zeta_2 \right) \right] \,,\label{dotp_xi_1}\\
\dot{p}_{\xi_2} &= \{ p_{\xi_2}, H_3 \} = -N \{ \int d^3\bx \mc{H}^3_0,
p_{\xi_2} \} \nn\\
&= -N \left\{ \frac{1}{\sqrt{g}}\left[ \frac{4\kappa^4 \alpha
U_0}{(1-2\kappa^2 \alpha U_0 \xi_2)^2}\left( p_{ij}p^{ij}-\frac{1}{3}p^2
\right) - \frac{p_{\zeta_2}^2}{6\alpha U_0 \xi_1^2} \right] +
\sqrt{g}\alpha\left( \zeta_2-U_0 R \right) \right\} \,.\label{dotp_xi_2}
\end{align}
Therefore we impose the following secondary constraints:
\begin{align}
\Phi^3_0 &= \int d^3\bx \mc{H}^3_0 \approx 0 \,,\qquad \Phi^3_i(\bx) =
\mc{H}^3_i \approx 0 \,,\qquad \Phi^3_4(\bx) = \zeta_2+\xi_1 \approx 0
\,,\label{Phi^3_(0-4)}\\
\Phi^3_5(\bx) &= \frac{1}{g} \left( pp_{\zeta_2} - \frac{1-2\kappa^2
\alpha U_0 \xi_2}{2\kappa^2 \alpha U_0 \xi_1} p_{\zeta_2}^2 \right) +
3\alpha^2 U_0 \xi_1^2 \left( \zeta_1 - 2U_0 D^i D_i \zeta_2 \right)
\approx 0 \,,\label{Phi^3_5}\\
\Phi^3_6(\bx) &= \frac{1}{g} \left( p_{ij}p^{ij} - \frac{1}{3} p^2 -
\frac{1}{6}\left(\frac{1-2\kappa^2 \alpha U_0 \xi_2}{2\kappa^2 \alpha
U_0 \xi_1}\right)^2 p_{\zeta_2}^2 \right) + \frac{\left(1-2\kappa^2
\alpha U_0 \xi_2\right)^2}{4\kappa^4 U_0} \left( \zeta_2-U_0 R \right)
\approx 0 \,.\label{Phi^3_6}
\end{align}
The superscript 3 in constraints $\mc{H}^3_0$, $\mc{H}^3_i$ and
$\Phi^3_n, n=0,1,\ldots,6$ refers to the value of the critical exponent
$z=3$ in the chosen model.
Note that in the right-hand side of expressions like \eqref{Phi^3_5} we
often omit the argument $\bx$ when there is no risk of confusion.
Evidently we are dealing with a more complicated structure of
constraints than in GR or in (projectable) HL gravity. We will see that
the constraints $\Phi^3_n(\bx), n=4,5,6$ are second-class and they
enable one to define the auxiliary fields $\xi_1, \zeta_1, \xi_2$ in
terms of the dynamical fields $g_{ij}, \zeta_2$ and their canonically
conjugated momenta $p^{ij}, p_{\zeta_2}$.

It is convenient to introduce a global smeared version of the momentum
constraints \eqref{momentum-constraints}:
\be \label{Phi^3_S}
\Phi^3_S(\chi^i) = \int d^3\bx \chi^i \mc{H}^3_i \,,
\ee
where $\chi^i$ ($i=1,2,3$) are arbitrary functions on $\Sigma_t$ which
vanish rapidly enough at infinity. As the name suggests, the momentum
constraints generate the infinitesimal spatial diffeomorphisms of the
dynamical variables:
\begin{align}
\{ \Phi^3_S(\chi^k), g_{ij} \} &= - \chi^k \p_k g_{ij} - g_{ik} \p_j
\chi^k - g_{jk} \p_i \chi^k \,,\nn\\
\{ \Phi^3_S(\chi^k), p^{ij} \} &= - \p_k \chi^k p^{ij} - \chi^k \p_k
p^{ij} + p^{ik} \p_k \chi^j + p^{jk}\p_k \chi^i \,,\nn\\
\{ \Phi^3_S(\chi^i), \zeta_2 \} &= - \chi^i \p_i \zeta_2 \,,\nn\\
\{ \Phi^3_S(\chi^i), p_{\zeta_2} \} &= - \p_i \chi^i p_{\zeta_2} -
\chi^i \p_i p_{\zeta_2} \,.\label{diffeom-generator}
\end{align}
Its PBs with the nondynamical fields and their canonically conjugated
momenta vanish. However, when $\xi_1, \zeta_1, \xi_2$ are solved in
terms of the dynamical variables, they behave as scalar fields, i.e.
like $\zeta_2$ in Eq.~\eqref{diffeom-generator}. The momenta are tensor
or scalar densities of weight $-1$ under spatial diffeomorphism, while
the fields are regular tensors or scalars.

In order to check that the secondary constraints $\Phi^3_I$
($I=0,1,2,\ldots,6$) are preserved under time evolution, we need to
evaluate their PBs with every constraint. First consider the momentum
constraint \eqref{Phi^3_S}. Its PB with the primary constraints
\eqref{pconstraints_z3} vanish. Since $\mc{H}^3_0$ is a scalar density
of weight $-1$ under spatial diffeomorphism, its integral
vanishes\footnotemark
\footnotetext{For a scalar density $\psi_w$ of weight $w$ we find: $\{
\Phi^3_S(\chi^i), \int d^3\bx\psi_w \} = - (w+1) \int d^3\bx \chi^i \p_i
\psi_w$.}
\be
\{ \Phi^3_S(\chi^i), \Phi^3_0 \} = 0 \,.
\ee
The PB of the momentum constraint with itself forms the Lie algebra
\be
\{ \Phi^3_S(\chi^i), \Phi^3_S(\psi^i) \} = \Phi^3_S(\chi^j \p_j \psi^i -
\psi^j \p_j \chi^i) \approx 0 \,,
\ee
Then consider the PBs with the rest of the secondary constraints in
Eqs~\eqref{Phi^3_(0-4)}--\eqref{Phi^3_6}. We see that the constraints
$\Phi^3_n(\bx), n=4,5,6$ have been defined to be scalars under spatial
diffeomorphism.

The Hamiltonian constraint $\Phi^3_0$ has vanishing PBs with $p_N$ and
$p_i$. The PBs with the rest of the primary constraints were calculated
in Eqs.~\eqref{dotp_zeta_1}--\eqref{dotp_xi_2}, and they vanish due to
the secondary constraints \eqref{Phi^3_(0-4)}--\eqref{Phi^3_6}.
Since the Hamiltonian constraint is global, its PB commutes with itself
\be
\{ \Phi^3_0, \Phi^3_0 \} = 0 \,.
\ee
Next consider the PBs with the secondary constraints $\Phi^3_n(\bx),
n=4,5,6$:
\be
\{ \Phi^3_4(\bx), \Phi^3_0 \} = - \frac{1}{\sqrt{g}}\frac{1}{3\alpha U_0
\xi_1} \left( p - \frac{1-2\kappa^2\alpha U_0\xi_2}{2\kappa^2\alpha
U_0\xi_1} p_{\zeta_2} \right) \,,
\ee
and the PBs of the last two secondary constraints turn out to be rather
complicated expressions:
\begin{multline} \label{Phi^3_5.Phi^3_0}
\{ \Phi^3_5(\bx), \Phi^3_0 \} = \frac{1}{g^{3/2}}
\frac{3\kappa^2}{1-2\kappa^2 \alpha U_0 \xi_2} \left[ p_{ij}p^{ij} -
\frac{1}{3}p^2 + \frac{1}{3}\left( \frac{1-2\kappa^2 \alpha U_0
\xi_2}{2\kappa^2 \alpha U_0 \xi_1}\right) p p_{\zeta_2} \right.\\
\left. - \frac{1}{2}\left( \frac{1-2\kappa^2 \alpha U_0 \xi_2}{2\kappa^2
\alpha U_0 \xi_1}\right)^2 p_{\zeta_2}^2 \right] p_{\zeta_2}
+ \frac{1}{\sqrt{g}} \left\{ \frac{1+2\kappa^2 \alpha U_0
\xi_2}{4\kappa^2} R \right.\\
- \alpha \left[ \frac{3}{2}\zeta_1 \zeta_2 + \xi_1 \left(
\frac{3}{2}\zeta_1-2U_0 D^i D_i \zeta_2 \right) + \frac{3}{2}\xi_2
\zeta_2 + U_0 \left( D_i \xi_1 D^i \zeta_2 - 2D^i D_i \xi_2 \right)
\right] p_{\zeta_2} \\
\left. +\alpha \left( p - 2\frac{1-2\kappa^2 \alpha U_0 \xi_2}{2\kappa^2
\alpha U_0 \xi_1} p_{\zeta_2} \right) \left( 2U_0 D^i D_i \xi_1 -
\zeta_1 - \xi_2 \right) \right\}\\
+ \frac{1}{\sqrt{g}} 6\alpha^2 U_0^2 \xi_1^2 \left[ \left( D^i D^j
\zeta_2 - D^{(i} \zeta_2 D^{j)} + \frac{1}{2} g^{ij} D^k D_k \right)
\right.\\
 \frac{2\kappa^2}{1-2\kappa^2 \alpha U_0 \xi_2} \left( 2p_{ij} -
\frac{g_{ij}}{3}\left( 2p + \frac{1-2\kappa^2 \alpha U_0
\xi_2}{2\kappa^2 \alpha U_0 \xi_1} p_{\zeta_2} \right) \right) \\
\left. + D^i D_i \frac{1}{3\alpha U_0 \xi_1} \left( p -
\frac{1-2\kappa^2\alpha U_0\xi_2}{2\kappa^2\alpha U_0\xi_1} p_{\zeta_2}
\right) \right]
\end{multline}
and
\begin{multline} \label{Phi^3_6.Phi^3_0}
\{ \Phi^3_6(\bx), \Phi^3_0 \} = \frac{1}{g^{3/2}}
\frac{p_{\zeta_2}}{\alpha U_0 \xi_1} \left[ p_{ij}p^{ij} -
\frac{1}{3}p^2 - \frac{1}{6}\left( \frac{1-2\kappa^2 \alpha U_0
\xi_2}{2\kappa^2 \alpha U_0 \xi_1}\right)^2 p_{\zeta_2} \right] \\
+ \frac{1}{\sqrt{g}} \left\{ \frac{1+2\kappa^2\alpha
U_0\xi_2}{2\kappa^2} \left( \frac{2}{3}Rp - R_{ij}p^{ij} \right) -
\alpha\xi_2\zeta_2 p \right.\\
+ \alpha U_0 \left[ 2\left( p^{ij}-\frac{1}{3}pg^{ij} \right) \left( D_i
D_j \xi_2 - 2D_{(i}\xi_1 D_{j)}\zeta_2 \right) - 2\xi_1 \left(
4p^{ij}-\frac{1}{3}pg^{ij} \right) D_i D_j \zeta_2 \right] \\
+ \frac{\alpha}{3} \left( \zeta_1 + \xi_2 - 2U_0 D^i D_i \xi_1 \right)
\left( \frac{1-2\kappa^2 \alpha U_0 \xi_2}{2\kappa^2 \alpha U_0
\xi_1}\right)^2 p_{\zeta_2} \\
- \frac{(1-2\kappa^2\alpha U_0\xi_2)^2}{4\kappa^4} \left[
\frac{1}{3\alpha U_0\xi_1}\left( p - \frac{1-2\kappa^2 \alpha U_0
\xi_2}{2\kappa^2 \alpha U_0 \xi_1} p_{\zeta_2} \right) \right.\\
\left.\left. - U_0 \left( R_{ij} - D_i D_j + g_{ij}D^k D_k \right)
\frac{2\kappa^2}{1-2\kappa^2 \alpha U_0 \xi_2}\left( 2p_{ij} -
\frac{g_{ij}}{3}\left( 2p + \frac{1-2\kappa^2 \alpha U_0
\xi_2}{2\kappa^2\alpha U_0\xi_1} p_{\zeta_2} \right) \right) \right]
\right\}
\end{multline}

The nonvanishing PBs between $\Phi^3_n(\bx), n=4,5,6$ and the primary
constraints \eqref{pconstraints_z3} are
\begin{align}
\{ \Phi^3_4(\bx), p_{\xi_1}(\by) \} &= \delta^3(\bx-\by) \,,\qquad \{
\Phi^3_5(\bx), p_{\zeta_1}(\by) \} = 3\alpha^2 U_0 \xi_1^2
\delta^3(\bx-\by) \,,\nn\\
\{ \Phi^3_5(\bx), p_{\xi_1}(\by) \} &= \left(
\frac{1}{g}\frac{1-2\kappa^2\alpha U_0 \xi_2}{2\kappa^2\alpha U_0 \xi_1}
p_{\zeta_2} + 6\alpha^2 U_0 \xi_1 \zeta_1 \right) \delta^3(\bx-\by)
\,,\nn\\
\{ \Phi^3_5(\bx), p_{\xi_2}(\by) \} &= \frac{1}{g}\frac{1}{\xi_1}
p_{\zeta_2}^2 \delta^3(\bx-\by) \,,\label{restPhiPBs.z3}\\
\{ \Phi^3_6(\bx), p_{\xi_1}(\by) \} &= \frac{1}{g}\frac{1}{3\xi_1}
\left( \frac{1-2\kappa^2 \alpha U_0 \xi_2}{2\kappa^2 \alpha U_0
\xi_1}\right)^2 p_{\zeta_2}^2 \delta^3(\bx-\by) \,,\nn\\
\{ \Phi^3_6(\bx), p_{\xi_2}(\by) \} &= \left(
\frac{1}{g}\frac{1}{3\xi_1} \frac{1-2\kappa^2 \alpha U_0
\xi_2}{2\kappa^2 \alpha U_0 \xi_1} p_{\zeta_2}^2 -
\frac{\alpha}{\kappa^2}\left(1-2\kappa^2\alpha U_0 \xi_2\right)
(\zeta_2-U_0 R) \right) \delta^3(\bx-\by) \,.\nn
\end{align}

Now we can consider the stability of the secondary constraints. First we
obtain that the Hamiltonian constraint and the momentum constraints are
preserved under time evolution:
\begin{align}
&\dot{\Phi}^3_0 = \{ \Phi^3_0, H_3 \} = \int d^3\bx \sqrt{g} \left(
\lambda_{\zeta_1} \alpha \Phi^3_4 + \lambda_{\xi_1} \frac{1}{3\alpha U_0
\xi_1^2} \Phi^3_5 + \lambda_{\xi_2} \frac{4\kappa^4 \alpha
U_0}{\left(1-2\kappa^2 \alpha U_0 \xi_2\right)^2} \Phi^3_6 \right)
\approx 0 \,,\nn\\
&\dot{\Phi}^3_S(\chi^i) = \{ \Phi^3_S(\chi^i), H_3 \} = \{
\Phi^3_S(\chi^i), \Phi^3_S(N^i) \} = \Phi^3_S(\chi^j \p_j N^i - N^j \p_j
\chi^i) \approx 0 \,.
\end{align}
The rest of the secondary constraints evolve under time as follows:
\begin{align}
\dot{\Phi}^3_4(\bx) &= \{ \Phi^3_4(\bx), H_3 \} = -
\frac{1}{\sqrt{g}}\frac{N}{3\alpha U_0 \xi_1} \left( p -
\frac{1-2\kappa^2\alpha U_0\xi_2}{2\kappa^2\alpha U_0\xi_1} p_{\zeta_2}
\right) + N^i \p_i \Phi^3_4 + \lambda_{\xi_1} \,,\nn\\
\dot{\Phi}^3_5(\bx) &= \{ \Phi^3_5(\bx), H_3 \} = N \{ \Phi^3_5(\bx),
\Phi^3_0 \} + N^i \p_i \Phi^3_5 + \lambda_{\zeta_1} 3\alpha^2 U_0
\xi_1^2 \nn\\
&+ \lambda_{\xi_1} \left( \frac{1}{g}\frac{1-2\kappa^2\alpha U_0
\xi_2}{2\kappa^2\alpha U_0 \xi_1} p_{\zeta_2} + 6\alpha^2 U_0 \xi_1
\zeta_1 \right) + \lambda_{\xi_2} \frac{1}{g}\frac{1}{\xi_1}
p_{\zeta_2}^2 \,,\label{dotPhi^3_(4-6)}\\
\dot{\Phi}^3_6(\bx) &= \{ \Phi^3_6(\bx), H_3 \} = N \{ \Phi^3_6(\bx),
\Phi^3_0 \} + N^i \p_i \Phi^3_6 + \lambda_{\xi_1}
\frac{1}{g}\frac{1}{3\xi_1} \left( \frac{1-2\kappa^2 \alpha U_0
\xi_2}{2\kappa^2 \alpha U_0 \xi_1}\right)^2 p_{\zeta_2}^2 \nn\\
&+ \lambda_{\xi_2} \left( \frac{1}{g}\frac{1}{3\xi_1} \frac{1-2\kappa^2
\alpha U_0 \xi_2}{2\kappa^2 \alpha U_0 \xi_1} p_{\zeta_2}^2 -
\frac{\alpha}{\kappa^2}\left(1-2\kappa^2\alpha U_0 \xi_2\right)
(\zeta_2-U_0 R) \right) \,,\nn
\end{align}
where Eqs.~\eqref{Phi^3_5.Phi^3_0} and \eqref{Phi^3_6.Phi^3_0} are
understood. We see that imposing $\dot{\Phi}^3_n(\bx)\approx 0, n=4,5,6$
in Eq.~\eqref{dotPhi^3_(4-6)} fixes the Lagrange multipliers
$\lambda_{\xi_1}$, $\lambda_{\zeta_1}$ and $\lambda_{\xi_2}$. Thus no
more constraints are required to fulfill the consistency conditions for
the secondary constraints. Hence we have obtained a stable constraint
surface in the phase space.

As noted above, we may use the constraints $\Phi^3_n(\bx), n=4,5,6$ to
eliminate the auxiliary fields $\zeta_1$, $\xi_1$ and $\xi_2$. To this
end, we can set the second-class constraints $p_{\zeta_1}$, $p_{\xi_1}$,
$p_{\xi_2}$ and $\Phi^3_n(\bx), n=4,5,6$ to vanish strongly by
introducing the Dirac bracket, and then solve the auxiliary fields in
terms of the dynamical variables from
Eqs.~\eqref{Phi^3_(0-4)}--\eqref{Phi^3_6}. This eliminates the fields
$\zeta_1, \xi_1, \xi_2$ and their canonically conjugated momenta from
the phase space of the system.

Let us then count the number of physical degrees of freedom (physical
d.o.f.) by using Dirac's formula:
\begin{multline}
\#(\text{physical d.o.f.}) = \frac{1}{2}\left[ \#(\text{canonical
variables}) - 2\times\#(\text{first-class constraints}) \right.\\
\left.- \#(\text{second-class constraints}) \right] \,.
\end{multline}
For nonpropagating zero modes we have 28 canonical variables ($N, N^i,
g_{ij}, \xi_1, \zeta_1, \xi_2,\zeta_2$, and their conjugated momenta), 8
first-class constraints ($p_N, p_i, \Phi^3_0, \mc{H}^3_i$), and 6
second-class constraints ($p_{\zeta_1}, p_{\xi_1}, p_{\xi_2}, \Phi^3_4,
\Phi^3_5, \Phi^3_6$). This yields
\be
\#(\text{nonpropagating physical d.o.f.}) = \frac{1}{2}(28-16-6) = 3
\ee
for the ($\bx$-independent) zero modes. For propagating modes we have 26
canonical variables ($N^i, g_{ij}, \xi_1, \zeta_1, \xi_2,\zeta_2$, and
their conjugated momenta), 6 first-class constraints ($p_i,
\mc{H}^3_i$), and 6 second-class constraints ($p_{\zeta_1}, p_{\xi_1},
p_{\xi_2}, \Phi^3_4, \Phi^3_5, \Phi^3_6$). Thus for propagating modes we
obtain
\be
\#(\text{propagating physical d.o.f.}) = \frac{1}{2}(26-12-6) = 4 \,.
\ee
For comparison, there are only two physical degrees of freedom in GR. As
another comparison, our analysis shows that CRG with the condition
$\phi=\phi(t)$ has one more physical degree of freedom than projectable
HL gravity, which has 2 zero modes and 3 propagating modes.
Interestingly, the number of physical modes is exactly the same as in
the modified $F(R)$ HL gravity
\cite{Chaichian:2010a,Carloni:2010,Chaichian:2010b}. 
One extra physical degree of freedom has its origin in the higher order
time derivatives present in the CRG action. The other extra propagating
mode is caused by the projectability condition similarly as in HL
gravity. 
Such extra degrees of freedom can be problematic since they may generate
extra (long range) forces that are not in agreement with observations.
One may be able to bring the number of physical degrees of freedom
closer to that of GR by introducing some extra gauge symmetry, which
generates some new constraints. 
Another possible way to deal with the extra propagating mode is to make
the scalar field $\zeta_2$ massive.

\subsection{The action \texorpdfstring{$S_4$}{S4}}
Since the action \eqref{Sz4.ADM.st.gf} is independent of the time
derivatives of $N$, $N^i$, $\eta$ and $\xi$, their canonically
conjugated momenta, $p_N$, $p_i$, $p_\eta$ and $p_\xi$ respectively, are
the primary constraints:
\be \label{pconstraints_z4}
p_N \approx 0 ,\qquad p_i(\bx) \approx 0 ,\qquad p_\eta(\bx) \approx 0
,\qquad p_\xi(\bx) \approx 0 .
\ee
The momenta conjugate to $g_{ij}$ and $\zeta$ are defined by
\begin{align}
p^{ij} &= \frac{\delta S_4}{\delta\dot{g}_{ij}} = \sqrt{g}\left(
\frac{1}{2\kappa^2}\mc{G}^{ijkl}K_{kl} + \alpha U_0 \left( 2\eta g^{ij}
\nabla_n \zeta + \xi \mc{G}^{ijkl}K_{kl} \right) \right)
\,,\label{p^ij_z4}\\
p_\zeta &= \frac{\delta S_4}{\delta\dot{\zeta}} = \sqrt{g} 4\alpha U_0
\eta K \,.\label{p_zeta}
\end{align}
The PB is postulated similarly as in Eq.~\eqref{PB_z3}.

Solving $\dot{\zeta}$ and $\dot{g}_{ij}$ in terms of the canonical
variables gives
\begin{align}
\dot{\zeta} &= N^i \p_i \zeta + \frac{N}{\sqrt{g}6\alpha U_0 \eta}
\left( p + \frac{1+2\kappa^2 \alpha U_0 \xi}{4\kappa^2 \alpha U_0 \eta}
p_\zeta \right) \,,\\
\dot{g}_{ij} &= 2NK_{ij} + 2D_{(i}N_{j)} \,,\\
K_{ij} &= \frac{1}{\sqrt{g}} \frac{1+2\kappa^2 \alpha U_0
\xi}{2\kappa^2} \mc{G}_{ijkl} \left[ p^{kl} + \frac{g^{kl}}{3}\left( p +
\frac{1+2\kappa^2 \alpha U_0 \xi}{4\kappa^2 \alpha U_0 \eta} p_\zeta
\right) \right]
\end{align}
The total Hamiltonian is obtained similarly as before:
\be
H_4 = \int d^3\bx \left( N\mc{H}^4_0 + N^i \mc{H}^4_i + \lambda_N p_N +
\lambda^i p_i + \lambda_\eta p_\eta + \lambda_\xi p_\xi \right) \,,
\ee
where the Hamiltonian and momentum constraints are defined by
\begin{align}
\mc{H}^4_0 &= \frac{1}{\sqrt{g}} \frac{2\kappa^2}{1+2\kappa^2 \alpha U_0
\xi} \left[ p_{ij}p^{ij} - \frac{1}{3}p^2 + \frac{1}{3}\left(
\frac{1+2\kappa^2 \alpha U_0 \xi}{4\kappa^2 \alpha U_0 \eta}\right) p
p_\zeta + \frac{1}{6}\left( \frac{1+2\kappa^2 \alpha U_0 \xi}{4\kappa^2
\alpha U_0 \eta}\right)^2 p_\zeta^2 \right] \nn\\
&- \sqrt{g}\left[ \frac{1-2\kappa^2 \alpha U_0 \xi}{2\kappa^2} R +
\alpha \left( \eta^2 - 4U_0 \eta D^i D_i \zeta + \xi\zeta \right)
\right] \,.\\
\mc{H}^4_i &= - 2g_{ij}D_k p^{jk} + D_i \zeta p_\zeta \nn\\
&= -2g_{ij}\p_k p^{jk} - \left( 2\p_j g_{ik} - \p_i
g_{jk} \right) p^{jk} + \p_i \zeta p_\zeta \,,
\end{align}

The primary constraints \eqref{pconstraints_z4} must be preserved under
the time evolution generated by the total Hamiltonian $H_4$ of the
system:
\begin{align}
\dot{p}_N &= \{ p_N, H_4 \} = -\int d^3\bx \mc{H}^4_0 \,,\\
\dot{p}_i &= \{ p_i, H_4 \} = -\mc{H}^4_i \,,\\
\dot{p}_\eta &= \{ p_\eta, H_4 \} = -N \{ \int d^3\bx \mc{H}^4_0, p_\eta
\} \nn\\
&= N \left[ \frac{1}{\sqrt{g}}\frac{1}{6\alpha U_0 \eta^2} \left(
pp_\zeta + \frac{1+2\kappa^2 \alpha U_0 \xi}{4\kappa^2 \alpha U_0 \eta}
p_\zeta^2 \right) + \sqrt{g}2\alpha \left( \eta - 2U_0 D^i D_i \zeta
\right) \right] \,,\\
\dot{p}_\xi &= \{ p_\xi, H_4 \} = -N \{ \int d^3\bx \mc{H}^4_0, p_\xi \}
\nn\\
&= N \left\{ \frac{1}{\sqrt{g}}\left[ \frac{4\kappa^4 \alpha
U_0}{(1+2\kappa^2 \alpha U_0 \xi)^2}\left( p_{ij}p^{ij}-\frac{1}{3}p^2
\right) - \frac{p_\zeta^2}{24\alpha U_0 \eta^2} \right] +
\sqrt{g}\alpha\left( \zeta-U_0 R \right) \right\} \,.
\end{align}
Therefore we impose the following secondary constraints:
\begin{align}
\Phi^4_0 &= \int d^3\bx \mc{H}^4_0 \approx 0 \,,\qquad \Phi^4_i(\bx) =
\mc{H}^4_i \approx 0 \,,\\
\Phi^4_4(\bx) &= \frac{1}{g} \left( pp_\zeta + \frac{1+2\kappa^2 \alpha
U_0 \xi}{4\kappa^2 \alpha U_0 \eta} p_\zeta^2 \right) + 12\alpha^2 U_0
\eta^2 \left( \eta - 2U_0 D^i D_i \zeta \right) \approx 0
\,,\label{Phi_4^4}\\
\Phi^4_5(\bx) &= \frac{1}{g} \left( p_{ij}p^{ij} - \frac{1}{3} p^2 -
\frac{1}{6}\left(\frac{1+2\kappa^2 \alpha U_0 \xi}{4\kappa^2 \alpha U_0
\eta}\right)^2 p_\zeta^2 \right) + \frac{\left(1+2\kappa^2 \alpha U_0
\xi\right)^2}{4\kappa^4 U_0} \left( \zeta-U_0 R \right) \approx 0
\,.\label{Phi_4^5}
\end{align}

The constraint structure is quite similar to the case $z=3$ we discussed
above. The only major difference is that there is one auxiliary field
less in the case $z=4$ than in the case $z=3$, and therefore there are
two second-class constraints less than in the case $z=3$ --- one primary
constraint and one secondary constraint. The consistency conditions
$\dot{\Phi}^4_n(\bx) \approx 0, n=4,5$ can be satisfied by fixing the
Lagrange multipliers $\lambda_\eta$ and $\lambda_\xi$. Thus we again
obtain a stable constraint surface. Once imposed, the second-class
constraints \eqref{Phi_4^4} and \eqref{Phi_4^5} define the auxiliary
fields $\eta$ and $\xi$ in terms of the dynamical variables. The number
of physical degrees of freedom is the same as in the case $z=3$.

\section{Conclusion and discussion}
\label{sec4}
Let us summarize our approach and results. Following the proposal of
Ref.~\cite{Nojiri:2010b}, we considered that the exotic perfect fluid of
CRG is generated by a scalar field $\phi$ that is constrained by
Eq.~\eqref{LMconstraint}. We obtained the ADM representations of such
CRG actions for the supposedly power-counting renormalizable $z=3$ and
super-renormalizable $z=4$ cases. The corresponding Lagrangians were
found to contain time derivatives of the dynamical fields up to fourth
order. In order to obtain Lagrangians with kinetic parts quadratic in
the first order time derivatives, we introduced some additional scalar
fields and took advantage of the constraint \eqref{LMconstraint} by
choosing $\p^\mu \phi$ to be parallel to the normal of the spatial
hypersurfaces.
The Hamiltonian formalism for the space-time decomposed actions was
developed. We showed that both actions have a consistent set of
constraints. All but one of the additional scalar fields turn out to be
auxiliary, which can be eliminated by imposing the second-class
constraints of the systems. The number of physical degrees of freedom
was found to be 3 for zero modes and 4 for propagating modes. Thus we
can conclude that the advantage of retaining general covariance and
local Lorentz invariance in the formulation of CRG comes with the price
of an extra physical degree of freedom --- compared to HL gravity. This
is a consequence of the higher order nature of the CRG action. Compared
to GR we found two extra propagating physical degrees of freedom. One
may be able to reduce the number of physical degrees of freedom closer
to that of GR by introducing some extra gauge symmetry or by making some
of the fields massive. These points, however, were not pursued further
in this paper.

It would be interesting to use the Pons formalism \cite{Pons:1989} for
the Hamiltonian analysis of CGR in order to gain deeper understanding of
these higher order derivative theories, and with the hope that a more
conclusive statement on their nature could be drawn. It would enable one
to analyze CRG without imposing the projectability condition on the
scalar field $\phi$, and consequently on the lapse $N$.
Another possible prospect is the application of perturbative constraints
to CRG. This could enable one to eliminate the possible problems caused
by the higher order time derivatives. It would also be interesting to
see whether the ``covariant vector gravity'' proposed in
Ref.~\cite{Nojiri:2010c} could provide even more fruitful results
compared to the present scalar formulation, since unlike $\p_\mu \phi$,
a vector field $A_\mu$ does not involve time derivatives.

\paragraph{Acknowledgements}
We wish to thank Sergei Odintsov for correspondence and invaluable
comments. 
The support of the Academy of Finland under the Projects No. 136539
and No. 140886 is gratefully acknowledged. M. O. is supported by the
Jenny and Antti Wihuri Foundation and by the Finnish Cultural
Foundation.

\end{document}